\documentclass[10pt,final,twocolumn]{IEEEtran}

\usepackage[noadjust]{cite}
\usepackage{psfrag}
\usepackage{latexsym, amsmath, color, amsfonts, amssymb,graphicx}
\usepackage{algorithm}
\usepackage{pdfsync}
\usepackage{booktabs}
\usepackage{algpseudocode}
\usepackage{amsmath}
\usepackage{graphics}
\usepackage{epsfig}
\usepackage{multirow}
\usepackage{subfigure}
\usepackage{tikz}
\usepackage{bm}

\newtheorem{theorem}{Theorem}[section]
\newtheorem{lemma}[theorem]{Lemma}

\newtheorem{proposition}[theorem]{Proposition}
\newtheorem{remark}[theorem]{Remark}

\newtheorem{example}[theorem]{Example}

\newtheorem{problem}[theorem]{Problem}

\newcommand{\hfs}{\hfill\ensuremath{\square}}

\begin{document}

\title{Fake-Acknowledgment Attack on ACK-based Sensor Power Schedule for Remote State Estimation}

\author{Yuzhe $\mathrm{Li}^{\star}$,  Daniel E.\ $\mathrm{Quevedo}^{\dag}$,~\IEEEmembership{Senior Member,~IEEE,}   Subhrakanti $\mathrm{Dey}^\diamond$,~\IEEEmembership{Senior Member,~IEEE,} and~Ling~$\mathrm{Shi}^{\star}$,~\IEEEmembership{Member,~IEEE}

\thanks{$\star$: Electronic and Computer Engineering, Hong Kong University of Science and Technology, Clear Water Bay, Kowloon, Hong Kong. Email: \{yliah, eesling\}@ust.hk.}
\thanks{$\dag$: Department of Electrical Engineering (EIM-E), The University of Paderborn, Germany, Email: dquevedo@ieee.org.}
\thanks{$\diamond$: Department of Engineering Sciences, Uppsala University, Sweden, Email: subhrakanti.dey@angstrom.uu.se.}}

%\thanks{A preliminary version of parts of this paper has been presented in XXX}

%The work by Y. Li and L. Shi was supported by an HK RGC GRF grant 618612.

\maketitle

\begin{abstract}
We consider a class of malicious attacks against remote state estimation. A sensor with limited resources adopts an acknowledgement (ACK)-based online power schedule to improve the remote state estimation performance. A malicious attacker can modify the ACKs from the remote estimator and convey fake information to the sensor. When the capability of the attacker is limited, we propose an attack strategy for the attacker and analyze the corresponding effect on the estimation performance. The possible responses of the sensor are studied and a condition for the sensor to discard ACKs and switch from online schedule to offline schedule is provided.
\end{abstract}


\section{Introduction}

Cyber-physical systems (CPS) have been a hot topic among both academic and industrial communities in the past few years. A wide application spectrum of CPS can be found in areas such as smart grid, intelligent transportation and environment monitoring~\cite{hespanha2007survey} with the integration of sensing, control, communication and computation. In most CPS infrastructures, wireless sensors are key components with advantages such as low cost, easy installation, self-power~\cite{gungor2009industrial}, when compared with traditional wired sensors. Therefore, wireless sensors have been increasingly equipped in CPS to replace wired sensors. The new issues due to their special characteristics have attracted much attention in recent years.

The first issue is how to allocate the energy of the sensor efficiently. Since most wireless sensors use on-board batteries which are difficult to replace, and are typically expected to work for several years without battery replacement, energy conservation is critical~\cite{quevedo2012kalman,imer2005optimal,leong2012power}. In~\cite{shi2011sensor}, the authors consider the sensor scheduling problem of whether it should send its data to a remote estimator or not due to the limited available communication energy. The optimal offline sensor schedule is derived, where the term ``offline'' means that the sensor designs its strategy before the process starting, i.e., independent of the state of the process. In~\cite{li2013online,han2014online,wu2012an}, the authors extend the result in~\cite{shi2011sensor} and propose a so-called ``online" sensor power schedules by utilizing the real-time information of the process. One typical category of online information is the ACK from the remote estimator, indicating whether the data packet from the sensor arrives successfully or not. The ACK-based online power schedule is proved to improve the estimation performance significantly~\cite{han2014online} compared to the offline schedule which only requires sending a $1$-bit ACK packet.

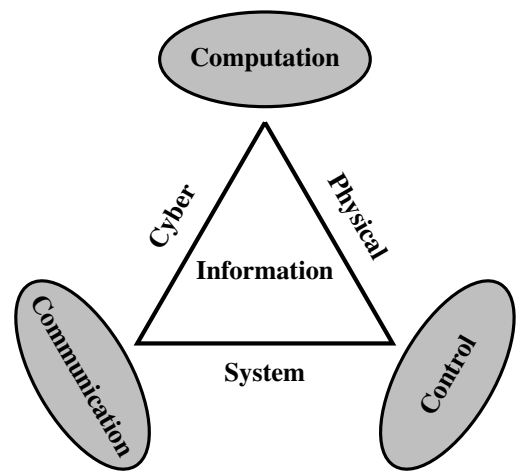
\begin{figure}[htp]

  \centering

\usetikzlibrary{shapes}
\begin{tikzpicture}[scale=0.7]

\draw[fill=lightgray, very thick, rotate around={60:(3.464,-2)}] (3.464,-2) ellipse (2 and 0.9);
\node[align=center, rotate=60] at (3.464,-2){\textbf{Control}};

\draw[fill=lightgray,very thick, rotate around={-60:(-3.464,-2)}] (-3.464,-2) ellipse (2 and 0.9);
\node[align=center, rotate=-60] at (-3.464,-2){\textbf{Communication}};

\draw[fill=lightgray,very thick] (0,4) ellipse (2 and 0.9);
\node[align=center] at (0,4){\textbf{Computation}};

\draw[ultra thick] (0,2.8)--(-2.425,-1.4)--(2.425,-1.4)--(0,2.8);

\node[align=center,rotate=60] at (-1.732,1){\textbf{Cyber}};
\node[align=center,rotate=-60] at (1.732,1){\textbf{Physical}};
\node[align=center] at (0,-2){\textbf{System}};
\node[align=center] at (0,0){\textbf{Information}};
\end{tikzpicture}

  \caption{Architecture of Cyber-Physical Systems.} \label{fig:cps_outline}

\end{figure}

However, as the communication is through a wireless channel, these sensors are more vulnerable to cyber security threats. In some safety-critical infrastructures, the wide use of CPS enlarges the risks and severities of such attacks. For example, as the largest and most complex CPS in the future, any severe attack on smart grid may have significant impacts on national economy and security or even loss of human life~\cite{mo2012cyber}. Therefore, the security issue is of fundamental importance to ensure the safe operation of CPS. There are two possible types of attacks that are commonly investigated in the literature: deception (integrity) attacks and denial-of-service (DoS) attacks~\cite{cardenas2008secure}, corresponding to the two traditional security goals \textit{integrity} and \textit{availability}. The integrity attacks on CPS focus on the integrity of data by modifying the data packet~\cite{liu2011false,mo2012integrity}, while the DoS attacks block the exchange of information including sensor measurements or control inputs between each part of CPS~\cite{li2013jamming,zhang2013optimal}. In practice, the integrity attacks studied in~\cite{liu2011false,mo2012integrity,bai2014kalman} often require comprehensive information about the system and modifications of the data. Such information is not needed in DoS attacks. In this work, we will consider both types of attacks.

Though the online power schedule in~\cite{han2014online} improves the estimation performance, the simple structure of the ACK packet makes it a more reachable (and likely) alternative for the adversary, under both integrity attacks and DoS attacks. Elucidating what is the possible attack patterns are and how they will affect the estimation performance are important to help us improve the design of CPS. This motivates us to investigate such potential class of malicious attacks against ACK-based remote state estimation.

The remainder of the work is organized as follows. Section II presents the system framework and states the main problem of interest. Section III provides some preliminaries about the optimal offline schedule and an online schedule. The analysis of our proposed fake-ACK attack pattern is given in Section IV. Numerical examples and simulations are demonstrated in Section V. Section VI provides some concluding remarks.

\textit{Notations}: $\mathbb{Z}$ denotes the set of all integers and $\mathbb{N}$ the positive integers. $\mathbb{R}$ is the set of real numbers. $\mathbb{R}^{n}$ is the $n$-dimensional Euclidean space. $\mathbb{S}_{+}^{n}$ (and $\mathbb{S}_{++}^{n}$) is the set of $n$ by $n$ positive semi-definite matrices (and positive definite matrices). When $X \in \mathbb{S}_{+}^{n}$ (and $\mathbb{S}_{++}^{n}$), we write $X \geqslant 0$ (and $X > 0$) and $X\geqslant Y$ if $X - Y \in \mathbb{S}_{+}^{n}$. The curled inequality symbols $\succeq$ and $\preceq$  (and their strict forms $\succ$ and $\prec$) are used to denote generalized componentwise inequalities between vectors: for vectors $\mathbf a=[a_1,a_2,...,a_n]', \mathbf b=[b_1,b_2,...,b_n]'$, we write $\mathbf a\succeq \mathbf b$ if $a_i\geqslant b_i$, for $i=1,2,...,n$. $\mathbf 1$ denotes vector with all entries one. $\mathrm{Tr}(\cdot)$ is the trace of a matrix. The superscript $'$ stands for transposition. For functions $g, h$ with appropriate domains, $g\circ h(x)$ stands for the function composition $g\big(h(x)\big)$, and $h^{n}(x) \triangleq h\big(h^{n-1}(x)\big)$, where $n\in\mathbb{N}$ and with $h^{0}(x) \triangleq x$. $\delta_{ij}$ is the Dirac delta function, i.e., $\delta_{ij}$ equals to $1$ when $i=j$, and $0$ otherwise. The notation $\mathbb{P}[\cdot]$ refers to probability and $\mathbb{E}[\cdot]$ to expectation.

\section{Problem Setup}

\subsection{System Model}

\begin{figure}[ht]
  \centering
  \includegraphics[width=8.5cm]{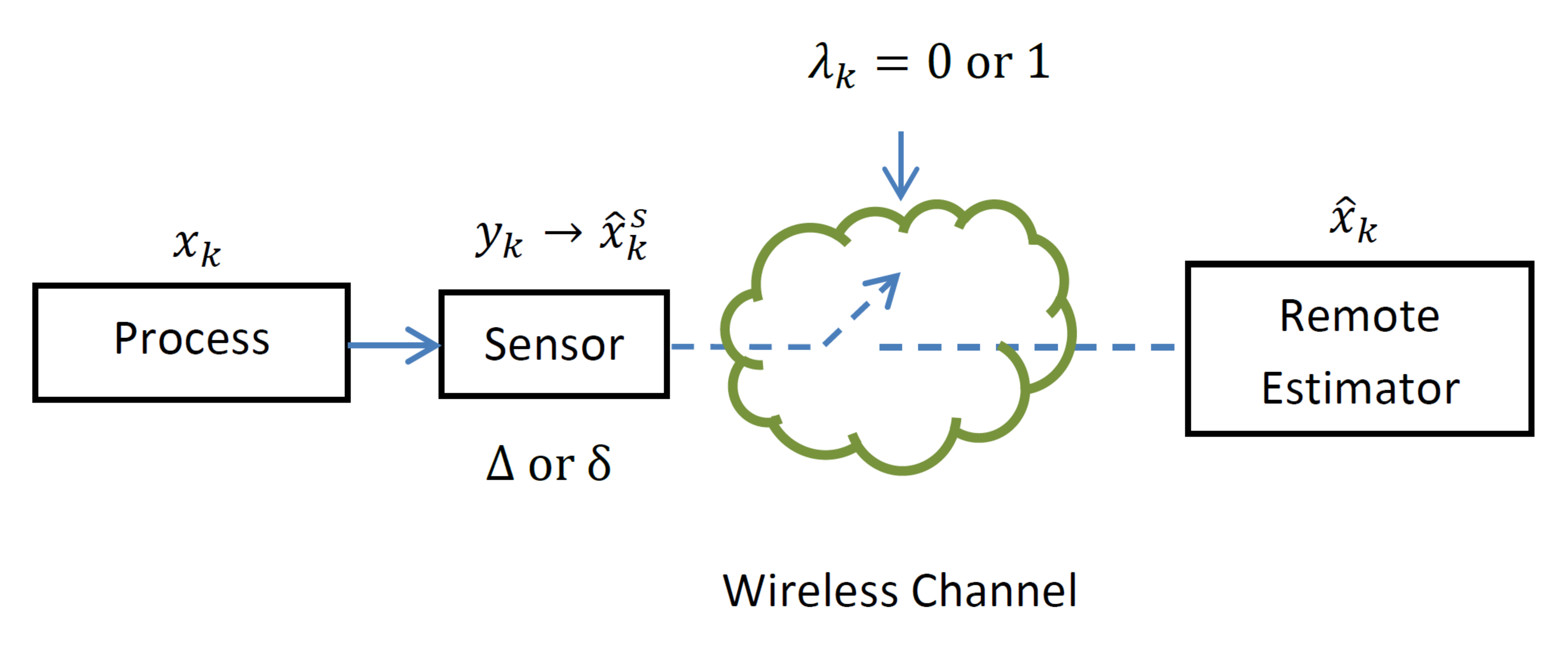}
  \caption{System Architecture.} \label{fig:system_normal}
\end{figure}

Our interest lies in the security of a remote state estimation as depicted in Fig.~\ref{fig:system_normal}. Here we consider a general discrete-time linear time-invariant (LTI) process of the form:
\begin{eqnarray}
  x_{k+1} & = & Ax_k + w_k, \label{eqn:process-dynamics} \\
  y_k & = & Cx_k + v_k,  \label{eqn:measurement-equation}
\end{eqnarray}
where  $k\in \mathbb{N}$, $x_k \in \mathbb{R}^{n_x}$ is the process state vector at time $k$, $y_k\in \mathbb{R}^{n_y}$ is the measurement taken by the sensor, $w_{k} \in\mathbb{R}^{n_x} $ and $v_k \in \mathbb{R}^{n_y}$ are zero-mean i.i.d.\ Gaussian noises with $\mathbb{E}[w_{k}w_{j}^\prime] =\delta_{kj}Q$ ($Q\geqslant 0$), $\mathbb{E}[v_{k}{v_{j}}^\prime] = \delta_{kj}R$ ($R > 0$), $\mathbb{E}[w_{k}{v_{j}}^\prime] = 0 \; \forall j,k\in\mathbb{N}$. The initial state $x_0$ is a zero-mean Gaussian random vector uncorrelated with $w_k$ and $v_k$ with covariance $\Pi_0\geqslant 0$. The pair $(A, C)$ is assumed to be observable and $(A, Q^{1/2})$ is controllable.

%\begin{figure}[htp]
%
%  \centering
%
%\usetikzlibrary{shapes}
%\begin{tikzpicture}[scale=1.4]
%\draw [thick] (0.2,0) rectangle (1.2,0.5);
%\node at (0.7,0.25){Process};
%\node at (0.7,0.7){$x_k$};
%\draw[very thick]  [->](1.2,0.25)--(1.6,0.25);
%\draw [thick] (1.6,0) rectangle (2.6,0.5);
%\node at (2.1,0.25){Sensor};
%\node at (2.1,0.7){$y_k\rightarrow\hat x_k^s$};
%\draw[ very thick][->] (2.6,0.25)--(2.9,0.25);
%
%\draw [thick,dashed] (2.9,-0.1) rectangle (4.7,0.6);
%\node[align=center] at (3.8,0.25){Communication\\Network};
%
%\draw[line width=4,purple] [->](3.75,1.4)--(3.75,0.8);
%\draw [thick,red] (3.25,1.55) rectangle (4.25,1.95);
%\node at (3.75,1.75)[red]{Attacker};
%\node at (4.5,1.2){DoS Attack};
%
%\draw[ very thick] [->](4.7,0.25)--(5,0.25);
%\draw [thick] (5,-0.1) rectangle (6.2,0.6);
%\node[align=center] at (5.6,0.25){Remote\\Estimator};
%\node at (5.6,0.8){$\hat x_k$};
%
%\end{tikzpicture}
%
%  \caption{The communication network is jammed by a
%    malicious attacker. This affects remote estimation performance.} \label{fig:system_outline}
%
%\end{figure}

We consider the so-called ``smart sensor" as described in~\cite{hovareshti2007sensor}, which first locally estimates the state $x_k$ based on all the measurements it has collected up to time $k$ and then transmits its local estimate to the remote estimator.

Denote $\hat{x}_k^s$ and $P_k^s$ as the sensor's local minimum mean-squared error (MMSE) estimate of the state $x_k$ and the corresponding error covariance:
\begin{eqnarray}
\hat{x}_k^s&=&\mathbb{E}[x_k|y_1,y_2,...,y_k], \label{eqn:state-all} \\
\hat{P}_k^s&=&\mathbb{E}[(x_k-\hat{x}_k^s)(x_k-\hat{x}_k^s)'|y_1,y_2,...,y_k], \label{eqn:error-covariance-all}
\end{eqnarray}
which can be calculated by a standard Kalman filter.

As the estimation error covariance of the Kalman filter converges to a unique value from any initial condition (\cite{sinopoli2004kalman}),  without loss of generality, we assume that the Kalman filter at the sensor side has entered the steady state and simplify our subsequent discussion by setting:
\begin{equation}\label{eqn:steady-state-assumption}
  P_k^s = \overline{P},~k \geqslant 1,
\end{equation}
where $\overline {P}$ is the steady-state error covariance. For notational ease, we define the Lyapunov and Riccati operators $h,\tilde{g}: \mathbb{S}_{+}^{n}\rightarrow \mathbb{S}_{+}^{n}$ as:
\begin{eqnarray*}
h(X) &\triangleq & AXA' + Q,\\
\tilde{g}(X) &\triangleq & X - XC'[CXC' + R]^{-1}CX.
\end{eqnarray*}
Then $\overline {P}$ is given by the unique positive semi-definite solution of $\tilde{g}\circ h(X) = X$ (see \cite{sinopoli2004kalman}).

The error covariance $\overline {P}$ has the following property:
\begin{lemma}(Lemma 2.3 in \cite{shi2011time}) \label{lemma:steady-state-error-property}
For $0\leqslant t_1\leqslant t_2$, the following inequality holds:
\begin{equation}
h^{t_1}(\overline{P}) \leqslant h^{t_2}(\overline{P}).
\end{equation}
In addition, if $t_1 < t_2$, then
\begin{equation} \label{eqn:steady-state-error-property}
\mathrm{Tr}\left(h^{t_1}(\overline{P})\right) < \mathrm{Tr}\left(h^{t_2}(\overline{P})\right).
\end{equation}
\end{lemma}

After obtaining $\hat{x}^s_k$, the sensor will transmit it as a data packet to the remote estimator. Due to fading and interference, random data drops will occur. As modelled in~\cite{shi2011sensor,han2014online}, we assume that the sensor has two transmission power levels: when using a higher energy $\Delta$, the packet will always arrive to the remote estimator; when using a lower energy $\delta$, the successful arrival rate is $\lambda\in(0,1)$. To simplify the following discussion, we denote the sensor power schedule as:
\begin{equation} \label{def:theta_gamma}
  \theta=\{\gamma_1,\gamma_2,...,\gamma_k,...\},
\end{equation}
where $\gamma_k=1$ or $0$ represents that the sensor chooses energy $\Delta$ or $\delta$, respectively, at time step $k$.

\subsection{Remote State Estimation}

The transmission of $\hat{x}^s_k$ between the sensor and the remote estimator can be characterized by a binary random process~$\{\lambda_k\}, k\in\mathbb{N}$:
\begin{equation*}
  \lambda_k =
  \begin{cases}
  1, &\text{if $\hat{x}^s_k$ arrives at time $k$,}\\
  0, & \text{otherwise (regarded as dropout).}
  \end{cases}
\end{equation*}
Denote $\hat{x}_k$ and $P_k$ as the remote estimator's own MMSE state estimate and the corresponding error covariance based on all the sensor data packets received up to time step $k$. From~\cite{li2013optimal} and~\cite{shi2010kalman}, the remote state estimate $\hat{x}_k$ obeys the recursion:
\begin{equation}\label{eqn:KF-remote-estimator}
\hat{x}_k = \left\{\begin{array}{ll}\hat{x}_{k}^s, &  \mathrm{if}~\lambda_k=1,\\
A\hat{x}_{k-1},  & \mathrm{if}~\lambda_k=0.
\end{array}\right.
\end{equation}
The corresponding state estimation error covariance $P_k$ satisfies:
\begin{equation}\label{eqn:KF-remote-error-covariance}
P_k = \left\{\begin{array}{ll}\overline{P}, &  \mathrm{if}~\lambda_k=1,\\
h(P_{k-1}),  & \mathrm{if}~\lambda_k=0.
\end{array}\right.
\end{equation}

Intuitively, higher transmission power leads to better estimation performance. However, in practice, the sensor has a limited energy budget, which motivates us to consider the following optimization problem:
\begin{problem} \label{problem:problem-1}
  \begin{eqnarray*}
    & & \min_{\theta}~~~J(\theta)\triangleq\limsup_{T\to\infty}\frac{1}{T}\sum_{k=1}^{T}\mathrm{Tr}\left(\mathbb{E}[P_k]\right),\\
    & & \mathrm{s.t.} ~~~\Phi(\theta)\triangleq\limsup_{T\to\infty}\frac{1}{T}\sum_{k=1}^{T}\mathbb{E}\left[\gamma_k\Delta+(1-\gamma_k)\delta\right]\leqslant\Psi,
  \end{eqnarray*}
  where  $\Psi$ is the average energy constraint for the sensor.
\end{problem}

\begin{remark}
  We assume that $\Delta$, $\delta$, and $\Psi$ are all positive rational numbers satisfying $\delta<\Psi<\Delta$.  \hfs
\end{remark}

\section{Preliminaries} \label{sec:preliminaries}

In this section, we will revise some preliminary results about the optimal offline schedule and an ACK-based online power schedule when there is no attacker.

\subsection{Optimal Offline Schedule}

When the sensor designs its power schedule without utilizing any online information, the optimal solution is provided by the following theorem:
\begin{theorem}[Optimal Offline Schedule Proposed in~\cite{shi2011sensor}] \label{thm:optimal_offline}
  Given the average energy constraint $\Psi$, suppose that $p$ and $q$ are two co-prime integers satisfying $\frac{p}{q}=\frac{\Psi-\delta}{\Delta-\delta}$. Then the optimal offline power schedule $\theta_{\text{off}}^\star$ to Problem~\ref{problem:problem-1} over a period $q$ can be constructed in the following form:
  \begin{equation*}
    \underbrace{(1\underbrace{0\cdots0}_{s_0+1\text{~times}})\cdots(1\underbrace{0\cdots0}_{s_0+1\text{times}})}_{n\text{~times}}\underbrace{(1\underbrace{0\cdots0}_{s_0\text{~times}})\cdots(1\underbrace{0\cdots0}_{s_0\text{times}})}_{m\text{~times}}
  \end{equation*}
  where $1$ and $0$ denotes the designed value for $\gamma_k$ in~\eqref{def:theta_gamma}, $m=q-p(s_0+1)$, $n=p(s_0+2)-q$, and $s_0$ is the largest integer such that $s_0\leqslant\frac{q}{p}-1$. \hfs
\end{theorem}

Under the optimal offline schedule $\theta_{\text{off}}^\star$, we have the following closed-form results about the corresponding energy consumption and estimation performance \cite{shi2011sensor}.

\begin{proposition}
  When $\theta_{\text{off}}^\star$ is used, the average energy cost is given by
  \begin{equation*}
    \Phi(\theta_{\text{off}}^\star)=\frac{(m+n)\Delta+(ms_0+ns_0+m)\delta}{ms_0+ns_0+2m+n}=\Psi,
  \end{equation*}
  and the trace of the expected average state estimation error covariance is given by:
  \begin{align*}
    J(\theta_{\text{off}}^\star)=&\frac{1}{m(s_0+2)+n(s_0+1)}\cdot\\
    &\mathrm{Tr}\Big\{m\big[1+\lambda(s_0+1)]\overline P+n\big[1+\lambda s_0)]\overline P\\
    &+m\sum_{i=1}^{s_0+1}\big[1+\lambda(s_0+1-i)\big](1-\lambda)^ih^i(\overline P)\\
    &+n\sum_{i=1}^{s_0}\big[1+\lambda(s_0-i)\big](1-\lambda)^ih^i(\overline P) \Big\}.
  \end{align*}  \hfs
\end{proposition}

\subsection{Online Power Schedule}

\begin{figure}[ht]
  \centering

  \includegraphics[width=8.5cm]{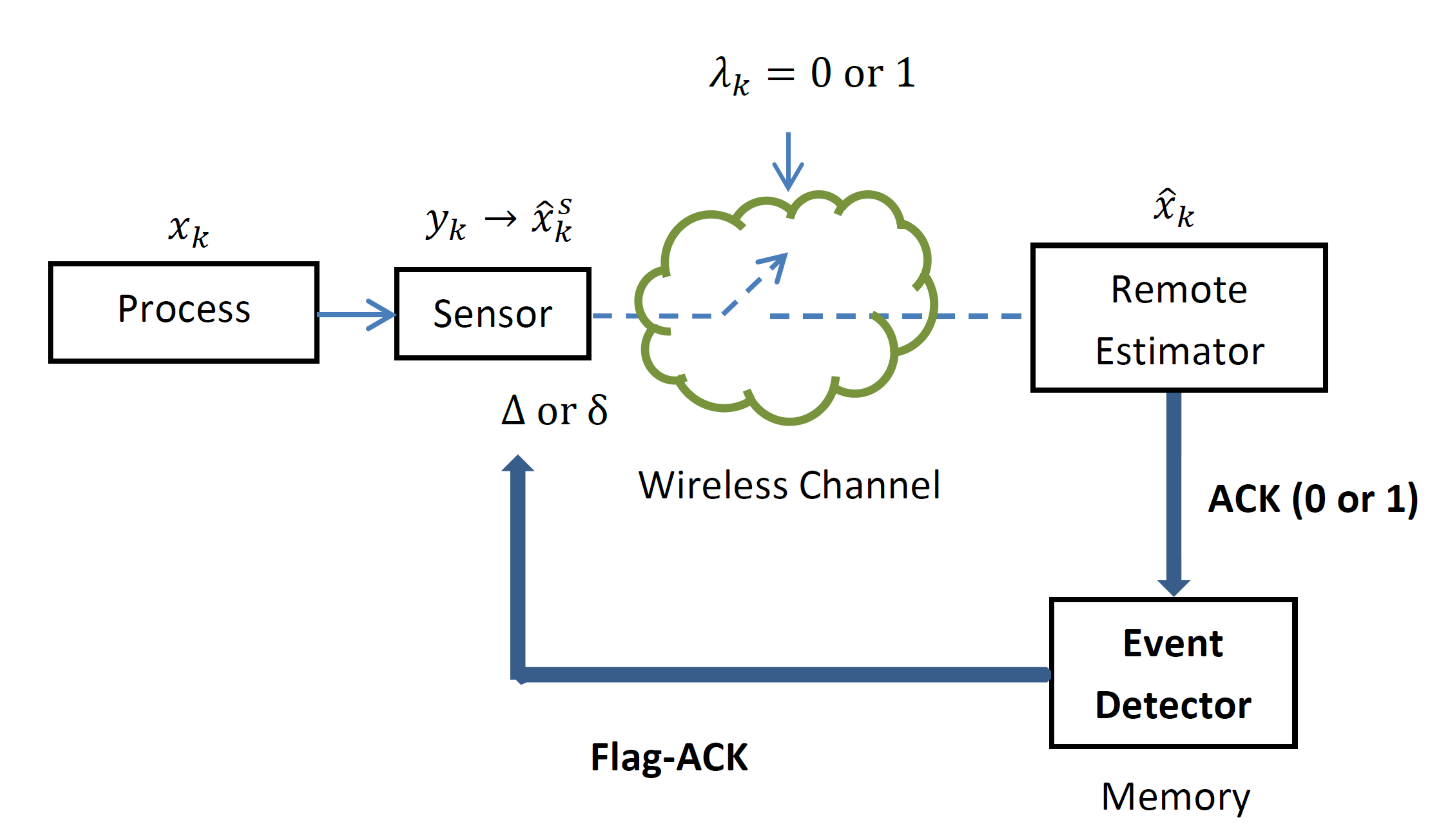}

  \caption{System Architecture under Online Sensor Schedule.} \label{fig:system_attack}
\end{figure}

The offline power schedules are designed before the system is running without using the realtime measurement or state information. In \cite{wu2012an}, the author showed that the utilization of the online information can improve the system performance while require less energy consumption. To further improve the estimation performance, an online power schedule based on the ACKs from the remote estimator was proposed in~\cite{han2014online} as follows: the remote estimator generates $1$-bit ACKs ($=\lambda_k$) to indicate whether the data packet arrives successfully or not and an event-detector at the remote estimator's side collects and stores the ACKs in an $L$-bit memory. Without loss of generality, assume the sensor uses $\Delta$ at the first time step. The memory is set to $11,...,11$ initially and the detector randomly chooses to activate $z_0$-bit memory ($z_0<L$) with probability $\mu$, or $(z_0+1)$-bit memory otherwise. At every time step, the memory shifts all bits one bit to the MSB direction with the existing MSB being deleted and the incoming ACK stored in the LSB. When the memory becomes $00,...,00$, the detector sends a flag-ACK to inform the sensor to use high power in the following time step. In the meanwhile, the memory is set to $11,...,11$ again (see Fig.~\ref{fig:system_attack}).

By properly designing the parameter $z_0$ and $\mu$, the average energy consumption can meet the constraint in Problem~\ref{problem:problem-1} and the corresponding estimation performance is given by the following theorem:
\begin{theorem} [Theorem 4.5 in \cite{han2014online}]
  Under the proposed online power schedule $\theta_{\text{on}}$, the trace of the average expected estimation error covariance is given by:
  \begin{align*}
    J(\theta_{\text{on}})=&\frac{\lambda}{1-\mu(1-\lambda)^{z_0+1}-\mu(1-\lambda)^{z_0+2}}\cdot\\
&\mathrm{Tr}\Big[\sum_{i=0}^{z_0}(1-\lambda)^ih^i(\overline P)+\mu(1-\lambda)^{z_0+1}h^{z_0+1}(\overline P)\Big].
  \end{align*} \hfs
\end{theorem}

Compared to the optimal offline schedule, the proposed online schedule $\theta_{\text{on}}$ has a better estimation performance under the same energy constraint, which is summarized in the following theorem~\cite{han2014online}:
\begin{theorem}[Theorem 5.3 in \cite{han2014online}]
  Under the same energy constraint as in Problem~\ref{problem:problem-1}, the proposed online schedule $\theta_{\text{on}}$ has a better estimation performance than the optimal offline schedule $\theta_{\text{off}}^\star$ provided in Theorem~\ref{thm:optimal_offline}, i.e.,
  \begin{equation*}
    J(\theta_{\text{on}})<J(\theta_{\text{off}}^\star).
  \end{equation*}
\end{theorem}

\section{Fake-acknowledgement Attack Against Remote Estimation}

As stated before, the ACK-based online power schedule can improve the estimation performance significantly compared to the offline schedule. However, the simple structure of the $1$-bit flag-ACK packet makes it a more reachable (and likely) alternative for the adversary, under both integrity attacks and DoS attacks. In this section, we propose a possible attack pattern for the attacker and investigate the corresponding effect on the estimation performance.

\subsection{Proposed Fake-ACK Attack Pattern}

Based on the communication protocol used by the event-detector, the attacker may adopt different types of attacks: when the flag-ACK is sent at every time step with value $1$ or $0$ (Type A), the attacker modifies the bit contained in the packet to launch the attack, i.e., the integrity attack; when the flag-ACK is sent only when the event happens (Type B), the attacker can simply adopt DoS attack to prevent the transmission of the packet. No matter what types of the communication protocols and attacks are used, the consequent results are the same: the time steps which would use energy $\Delta$ now use $\delta$ instead under the attack. Therefore, without loss of generality, we will assume that the event-detector adopts protocol Type B and the attacker launches DoS attack in the following discussion (the analysis of the other case is almost the same).

Assume that the attacker also has an energy constraint and cannot launch attack at every time step. Otherwise, it is straightforward to see that the attacker will attack all the flag-ACKs with value $1$ to prevent the transmission. Let the proportion of the flag-ACK packets that can be prevented by the attacker among all the flag-ACK packets be constrained to be $\beta\in(0,1)$ and $t, r(t>r)$ are two co-prime integers satisfying $\frac{r}{t}=\beta$. In~\cite{zhang2013optimal}, the optimal DoS attack pattern is proved to be launching the attack consecutively so that the effect on the estimation performance is maximized. This motivates us to consider the specific attack pattern of interest described as follows:
\begin{enumerate}
  \item the attacker follows a periodic pattern;
  \item the attacker maintains a counter and launches DoS attacks to prevent the transmission of the first $r$ coming flag-ACKs, and leave the rest $t-r$ coming flag-ACK with value $1$ unchanged;
  \item after every $t$ coming flag-ACKs, the counter is set to $0$ again.
\end{enumerate}
This is summarized in Algorithm~\ref{algorithm:attack_pattern}.

\begin{algorithm} [h]
\caption{Attacking Pattern for The Attacker} \label{algorithm:attack_pattern}
\begin{algorithmic}[1]
\State Process begins;
\State $counter=0$;
\While{1}
\If {a flag-ACK packet is sent}
\If{$counter\leqslant r$}
\State block the transmission of the flag-ACK packet;
\Else
\State do not block;
\EndIf
\State $counter=counter+1$;
\EndIf
\If{$counter==t$}
\State $counter=0$;
\EndIf
\EndWhile
\end{algorithmic}
\end{algorithm}

\begin{figure}[ht]
  \centering

  \includegraphics[width=8cm]{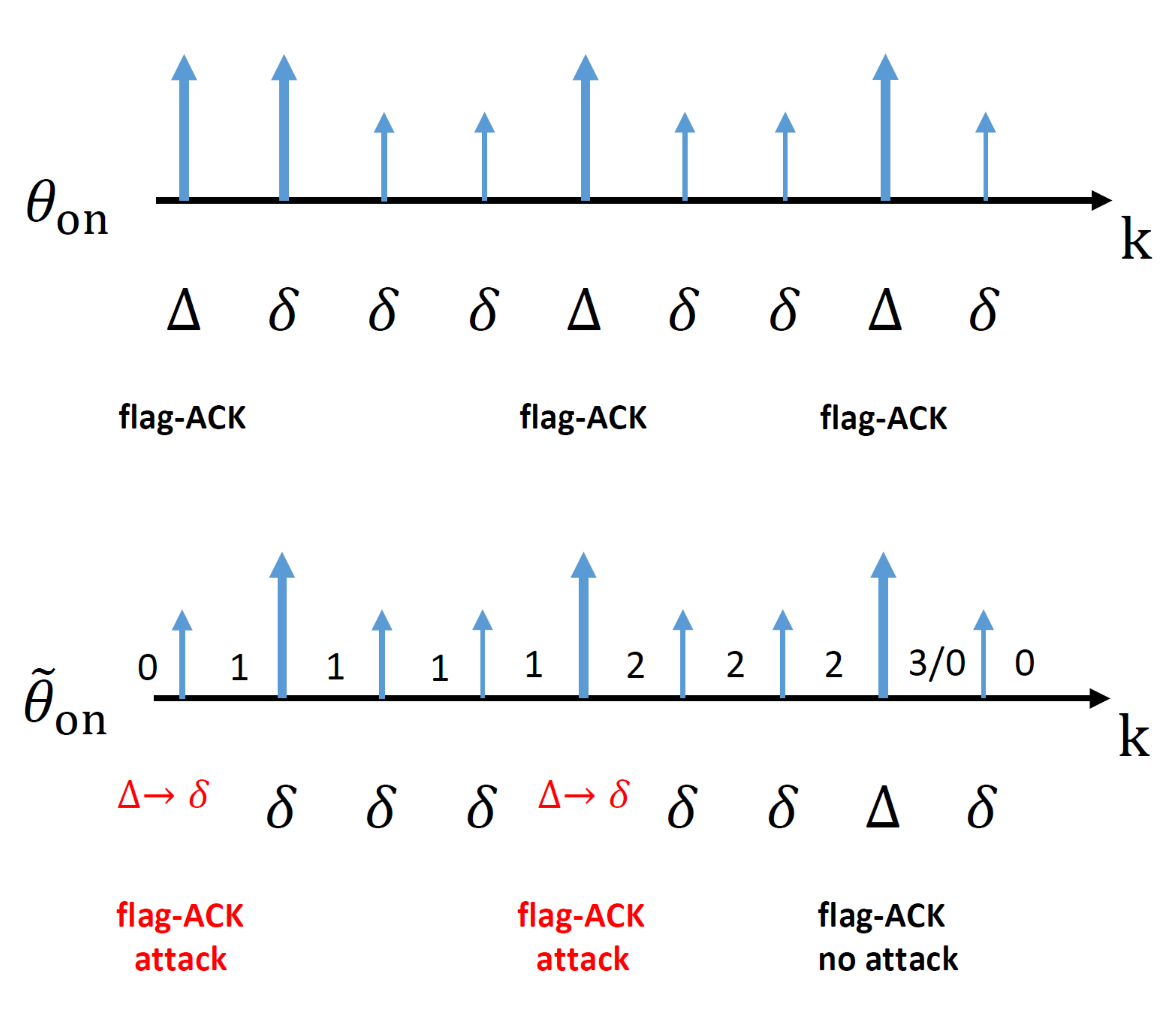}

  \caption{Comparison between $\theta_{\text{on}}$ and $\Tilde\theta_{\text{on}}$} \label{fig:attack_example}
\end{figure}

\begin{example}
  We use a simple example to illustrate how this attack pattern works. Assume that $z_0=2$, $\beta=\frac{2}{3}$ (thus $r=2$ and $t=3$), one possible realization and comparison between the online schedule $\theta_{\text{on}}$ and the one under attack $\Tilde\theta_{\text{on}}$ are shown in Fig.~\ref{fig:attack_example} (the long arrows represent the arrival packets while the short one is for the dropped packets; the number between arrows is the value of the counter). In the realization of $\theta_{\text{on}}$ without attack, the second and the third flag-ACK are sent due to two ($z_0=2$) consecutive packet losses. When there is an attacker following the pattern in Algorithm~\ref{algorithm:attack_pattern}, it will attack the first two ($r=2$) flag-ACKs, let the third one transmitted, and reset the counter for a new period.
\end{example}

\subsection{Performance Analysis}

Denote the estimation performance of $\theta_{\text{on}}$ under the proposed fake-ACK attack as $J(\Tilde\theta_{\text{on}})$. We will investigate the analytical form of $J(\Tilde\theta_{\text{on}})$ in the following part.

Note that due to the recursion of the dynamics in~\eqref{eqn:KF-remote-error-covariance}, the covariance $P_k$ can only take value in the infinitely countable set $\{\overline P, h(\overline P), h^2(\overline P),...\}$. Denote $\tau_k\in\mathbb Z$ as the holding time since the most recent time when the remote estimator successfully received the data from sensor:
\begin{equation} \label{definiton_tau}
  \tau_k\triangleq k-\max_{1\leqslant t \leqslant k}\{t : \lambda_t=1\}.
\end{equation}
Then we have
\begin{equation*}
P_k = \left\{\begin{array}{ll}\overline{P}, &  \mathrm{if}~\lambda_k=1,\\
h^{\tau_k}(\overline P),  & \mathrm{if}~\lambda_k=0.
\end{array}\right.
\end{equation*}

Define the value of the attacker's counter at time step $k$ as $\sigma_k\in\{0,1,2,...,t-1\}$ and the state of the process at time step $k$ as:
\begin{equation*}
  S_k={(\tau_k,\sigma_k)}.
\end{equation*}

As the strategy of the attacker guarantees that the flag-ACK will be sent to the sensor when the value of the counter is larger than $r$, the covariance $P_k$ will be bounded and $\tau_k$ only takes value in a finite set. Therefore, based on the recursion in~\eqref{eqn:KF-remote-error-covariance}, it is easy to verify that the process $\{S_k\}$ constitutes a finite state stationary Markov chain.

\begin{remark}
  Note that the design of the parameter $\mu$ in the online sensor schedule $\theta_{\text{on}}$ is to tune the activated memory length between $z_0$ and $z_0+1$ in the event-detector to satisfy the energy constraint. When the energy constraint is in some simple form, e.g., $\frac{1}{2}$ or $\frac{1}{3}$, a fixed length activated memory with $z_0$ is sufficient. Without loss of generality, we assume that the event-detector adopts a fixed activated memory with length $z_0$ to facilitate the discussion and simplify the notation below. We will later show that our results also apply to the general case with variant length. \hfs
\end{remark}

According to the mechanism of the online sensor schedule $\theta_{\text{on}}$ and the attack strategy, $\tau_k$ only takes values in the set $\{0,1,2,...,z_0(\sigma_k+1)\}$ when $\sigma_k\leqslant r$ and in the set $\{0,1,2,...,z_0\}$ when $r<\sigma_k<t$. Then the state set can be constructed as:
\begin{align*}
  \mathcal S\triangleq\Big\{&(0,0),(1,0),..,(z_0-1,0),(z_0,0),\\
  &(0,1),(1,1),...,(2z_0-1,1),(2z_0,1),\\
  &...\\
  &(0,r),(1,r),...,\big(z_0(r+1)-1,r\big),\big(z_0(r+1),r\big),\\
  &(0,r+1),(1,r+1),..,(z_0-1,r+1),(z_0,r+1),\\
  &(0,r+2),(1,r+2),..,(z_0-1,r+2),(z_0,r+2),\\
  &...\\
  &(0,t-1),(1,t-1),..,(z_0-1,t-1),(z_0,t-1)\Big\}.
\end{align*}

The total number of states can be calculated as:
\begin{align*}
|\mathcal S|&=\sum_{i=0}^r\big[z_0(i+1)+1\big]+\big[(t-1)-(r+1)+1\big](z_0+1)\\
  &=\frac{1}{2}r(r+1)z_0+t(z_0+1).
\end{align*}

It is convenient to label each state $\epsilon_i$ in $\mathcal S$ in a sequential order from $i=1$ to $|\mathcal S|$. Then we can define the state transition probability matrix as:
\begin{equation*}
  \mathcal T_s=\{T_{ij}\}_{|\mathcal S|\times |\mathcal S|},
\end{equation*}
where
\begin{equation*}
 T_{ij}\triangleq \mathbb P[S_{k+1}=\epsilon_j|S_k=\epsilon_i].
\end{equation*}

Based on the mechanism of the online sensor schedule $\theta_{\text{on}}$ and the attack strategy, we can derive the transition probability from $S_{k+1}=(\tau_{k+1},\sigma_k)$ to $S_k=(\tau_{k},\sigma_k)$ as follows:

$1)$ when $\tau_k\mod z_0\neq0$ or $\tau_k=0$, the event detector will not send the flag-ACK (and no attack as a consequence). Thus:
\begin{align*}
  &\mathbb P\big[S_{k+1}|S_k\big]=\begin{cases}
  \lambda, &\text{if~} (\tau_{k+1},\sigma_{k+1})=(0,\sigma_k),\\
  1-\lambda, & \text{if~} (\tau_{k+1},\sigma_{k+1})=(\tau_k+1,\sigma_k),\\
  0,&\text{otherwise};
  \end{cases}
\end{align*}

$2)$ when $\tau_k\mod z_0=0$, $\tau_k\neq0$ and $\sigma_k<r$, the memory satisfies the triggering condition, the flag-ACK will be sent and the attack will be launched with the counter increasing by $1$. Thus:
\begin{align*}
  &\mathbb P\big[S_{k+1}|S_k\big]=\begin{cases}
  \lambda, &\text{if~} (\tau_{k+1},\sigma_{k+1})=(0,\sigma_k+1),\\
  1-\lambda, & \text{if~} (\tau_{k+1},\sigma_{k+1})=(\tau_k+1,\sigma_k+1),\\
  0,&\text{otherwise};
  \end{cases}
\end{align*}

$3)$ when $\tau_k\mod z_0=0$, $\tau_k\neq0$ and $r\leqslant\sigma_k<t-1$, the memory satisfies the triggering condition, the flag-ACK will be sent without being attacked and the counter increasing by $1$. Thus:
\begin{align*}
  &\mathbb P\big[S_{k+1}|S_k\big]=\begin{cases}
  1, &\text{if~} (\tau_{k+1},\sigma_{k+1})=(0,\sigma_k+1),\\
  0,&\text{otherwise};
  \end{cases}
\end{align*}

$4)$ when $\tau_k\mod z_0=0$, $\tau_k\neq0$ and $\sigma_k=t-1$, the memory satisfies the triggering condition, the flag-ACK will be sent without being attacked and the counter is reset to $0$. Thus:
\begin{align*}
  &\mathbb P\big[S_{k+1}|S_k\big]=\begin{cases}
  1, &\text{if~} (\tau_{k+1},\sigma_{k+1})=(0,0),\\
  0,&\text{otherwise}.
  \end{cases}
\end{align*}

Based on the above discussion, we can obtain the expression of the transition matrix $\mathcal T_s$. Due to space limitation, we provide a simple example of $\mathcal T_s$ when $z_0=2$, $r=1$ and $t=3$ (therefore $|\mathcal S|=11$) in~\eqref{eqn:table}.

\begin{figure*}[t]
\begin{equation}\label{eqn:table}
\mathcal T_s=\left[
  \begin{array}{ccccccccccc}
     \bm \lambda & \bm\lambda  &\bm\lambda  &0 &0 &0 &0&0 &0&0&\bm1\\
    \bm{1-\lambda} & 0  & 0 &0 &0 &0 &0&0 & 0 &0&0\\
    0 & \bm{1-\lambda}  & 0 &0 &0 &0 &0&0 &0&0&0\\
    0 &0   & 0  &\bm\lambda &\bm\lambda &0 &\bm\lambda&0 &0&0&0\\
    0 &0   & 0  &\bm{1-\lambda} &0 &0 &0&0 &0&0&0\\
    0 &0   &0   &0 &\bm{1-\lambda}  &0 &0 &0 &0&0&0\\
    0 &0   &\bm{1-\lambda}  &0&0 &0 &0&0 &0&0&0\\
    0 &0   &0   &0 &0 &0 &\bm{1-\lambda}&0 &0&0&0\\
    0 &0   &0   &0 &0 &\bm1 &0&\bm1 &\bm\lambda &\bm\lambda &0\\
    0 &0   &0   &0 &0 &0 &0&0 &\bm{1-\lambda}& 0&0\\
    0 &0   &0   &0 &0 &0 &0&0 & 0&\bm{1-\lambda}&0
  \end{array}
\right]
\end{equation}
\end{figure*}

Based on the well-established ergodic Markov chain theory~\cite{norris1998markov}, $\{S_k\}$ has a stationary probability distribution of states, which is denoted as:
\begin{equation*}
  \Pi=\Big[\mathbb P[\epsilon_1],\mathbb P[\epsilon_2],...,\mathbb P[\epsilon_{|\mathcal S|}]\Big]',
\end{equation*}

When $\{S_k\}$ enters the steady state, we have
\begin{equation} \label{eqn:steady_markov_condition}
  \begin{cases}
  &\mathcal T_s\Pi=\Pi,\\
  &\sum_{i=1}^{|\mathcal S|}\Pi(i)=1,
  \end{cases}
\end{equation}
where $\Pi(i)$ is the $i$-th element of $\Pi$.

By solving~\eqref{eqn:steady_markov_condition}, we can obtain the stationary distribution of state, denoted as $\Pi^\star$. Then we can calculate the estimation performance of $\theta_{\text{on}}$ under the proposed fake-ACK attack as follows:
\begin{align*}
    J(\Tilde\theta_{\text{on}})=&\limsup_{T\to\infty}\frac{1}{T}\sum_{k=1}^{T}\mathrm{Tr}\left(\mathbb{E}[P_k]\right)\\
=&\limsup_{k\to\infty}\mathrm{Tr}\left(\mathbb{E}[P_k]\right)\\
=&\sum_{i=1}^{|\mathcal S|}\Pi^\star(i)\mathrm{Tr}\Big(\mathbb{E}\big[P_k(\epsilon_i)\big]\Big).
\end{align*}

\subsection{Discussion}

To analyze the effect of the proposed attacking pattern on $\theta_{\text{on}}$, we first consider two extreme cases.

When the energy constraint for the attacker is $\beta=1$, the attacker can block the transmissions of all the flag-ACKs, resulting in the sensor using the low power $\delta$ within the entire time-horizon. The corresponding estimation performance can be calculated as:
\begin{align*}
    J(\Tilde\theta_{\text{on}})_{\text{max}}=&\limsup_{T\to\infty}\frac{1}{T}\sum_{k=1}^{T}\mathrm{Tr}\left(\mathbb{E}[P_k]\right)\\
=&\limsup_{k\to\infty}\mathrm{Tr}\left(\mathbb{E}[P_k]\right)\\
=&\sum_{i=0}^{+\infty}\lambda(1-\lambda)^i\mathrm{Tr}\big(h^i(\overline P)\big).
\end{align*}
Note that the necessary and sufficient condition for the stability of the above case is simply given by $\rho(A)(1-\lambda)<1$, where $\rho(A)$ is the maximum eigenvalue of $A$. Since, as proved in~\cite{li2013optimal} and~\cite{shi2011sensor}, more energy budget is always beneficial, we have:
\begin{equation} \label{eqn:J_max_attack}
J(\Tilde\theta_{\text{on}})_{\text{max}}=J(\theta_{\text{off}}^\star)_{(n=0)}\geqslant J(\theta_{\text{off}}^\star),
\end{equation}
where $n$ is the number of high energy $\delta$ that can be used within each period of $J(\theta_{\text{off}}^\star)$ defined in Theorem~\ref{thm:optimal_offline}.

One the other hand, when $\beta=0$, the attacker cannot block any flag-ACK and the online sensor schedule will operate normally. Therefore, we have:
\begin{equation} \label{eqn:J_min_attack}
J(\Tilde\theta_{\text{on}})_{\text{min}}=J(\theta_{\text{on}}).
\end{equation}

Combining \eqref{eqn:J_max_attack} and \eqref{eqn:J_min_attack}, we have:
\begin{equation*}
J(\theta_{\text{on}})=J(\Tilde\theta_{\text{on}})_{\text{min}} \leqslant J(\theta_{\text{off}}^\star)\leqslant J(\Tilde\theta_{\text{on}})_{\text{max}}.
\end{equation*}

As a consequence, when the capability of the attacker is limited, keeping the online schedule will still guarantee a better performance than the offline schedule $\theta_{\text{off}}^\star$. Otherwise, when the attacker has sufficient energy budget, the estimation performance under the online sensor schedule $\theta_{\text{on}}$, $J(\Tilde\theta_{\text{on}})$, will be even worse than the offline schedule $\theta_{\text{off}}^\star$. In such case, not using ACK is preferable to using fake ACK and the reasonable choice for the sensor is to adopt the offline schedule rather than keeping the online schedule. Furthermore, as the performance metric $J(\Tilde\theta_{\text{on}})$ is growing with the energy budget $\beta$, there exists a threshold value $\overline \beta$ such that (as later shown in Fig.~\ref{fig:simulation_example2}):
\begin{equation*}
J(\Tilde\theta_{\text{on}})_{\beta=\overline\beta}=J(\theta_{\text{off}}^\star).
\end{equation*}
Hence, in practice, based on the information about the attacker's capability, the sensor can choose different types of schedules to obtain a better performance, i.e.,
\begin{equation*}
  \Tilde\theta_{\text{on}}=\begin{cases}
  \theta_{\text{on}}, &\text{when~}\beta<\overline \beta,\\
  \theta_{\text{off}}^\star, &\text{when~}\beta\geqslant\overline \beta,
  \end{cases}
\end{equation*}
where $\overline \beta$ can be calculated numerically. How to obtain the value of $\overline \beta$ analytically is an interesting problem which will be considered in our future work.

\begin{remark}
  Note that the sensor can detect the existence of the attacker by checking the arrival rate of flag-ACKs based on the theoretical arrival rate provided in~\cite{han2014online}. The observation will enable the sensor to estimate the capability of the attacker. On the other hand, when the sensor is aware of the existence of the attacker, the attacker will also need to re-design his strategy to avoid being detected, which will result in a dynamic decision-making process between both sides. We will investigate this problem under a game-theoretic framework in the future. \hfs
\end{remark}

\section{Simulation}

In this section, we provide numerical examples to illustrate our results in different situations.

Define
\begin{equation*}
  J_k(\theta)=\frac{1}{k}\sum_{i=1}^{k}\mathrm{Tr} \left(\mathbb{E}[P_i]\right),
\end{equation*}
as the empirical approximation (via 100000 Monte Carlo simulations) of $J(\theta)$ at every time instant $k$.

\begin{figure}[ht]
  \centering

  \includegraphics[width=8.5cm]{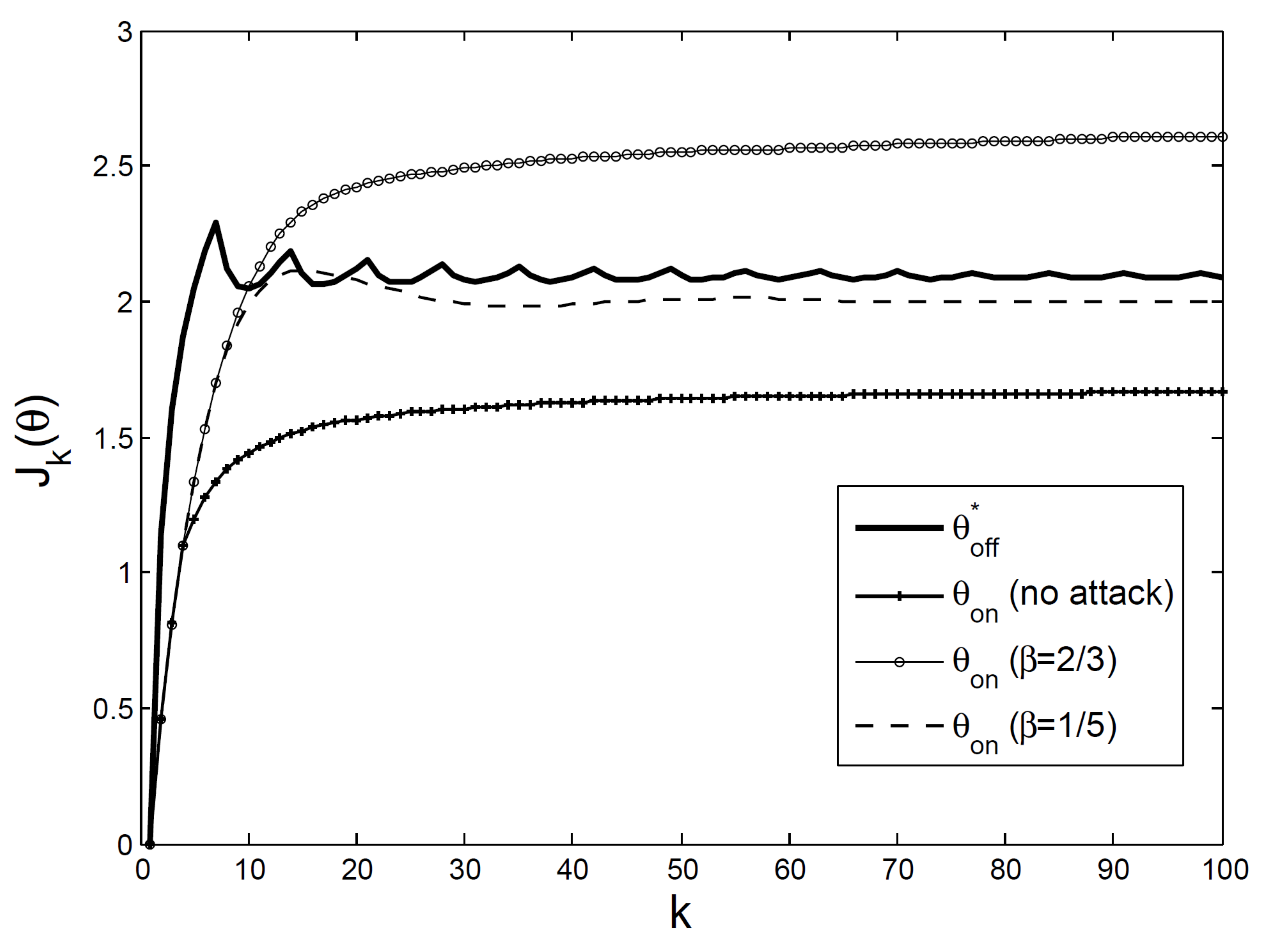}

  \caption{Comparison between $J(\theta^\star_{\text{off}})$, $J(\theta_{\text{on}})$ and $J(\Tilde\theta_{\text{on}})$.} \label{fig:simulation_example1}
\end{figure}

\begin{figure}[ht]
  \centering

  \includegraphics[width=8.5cm]{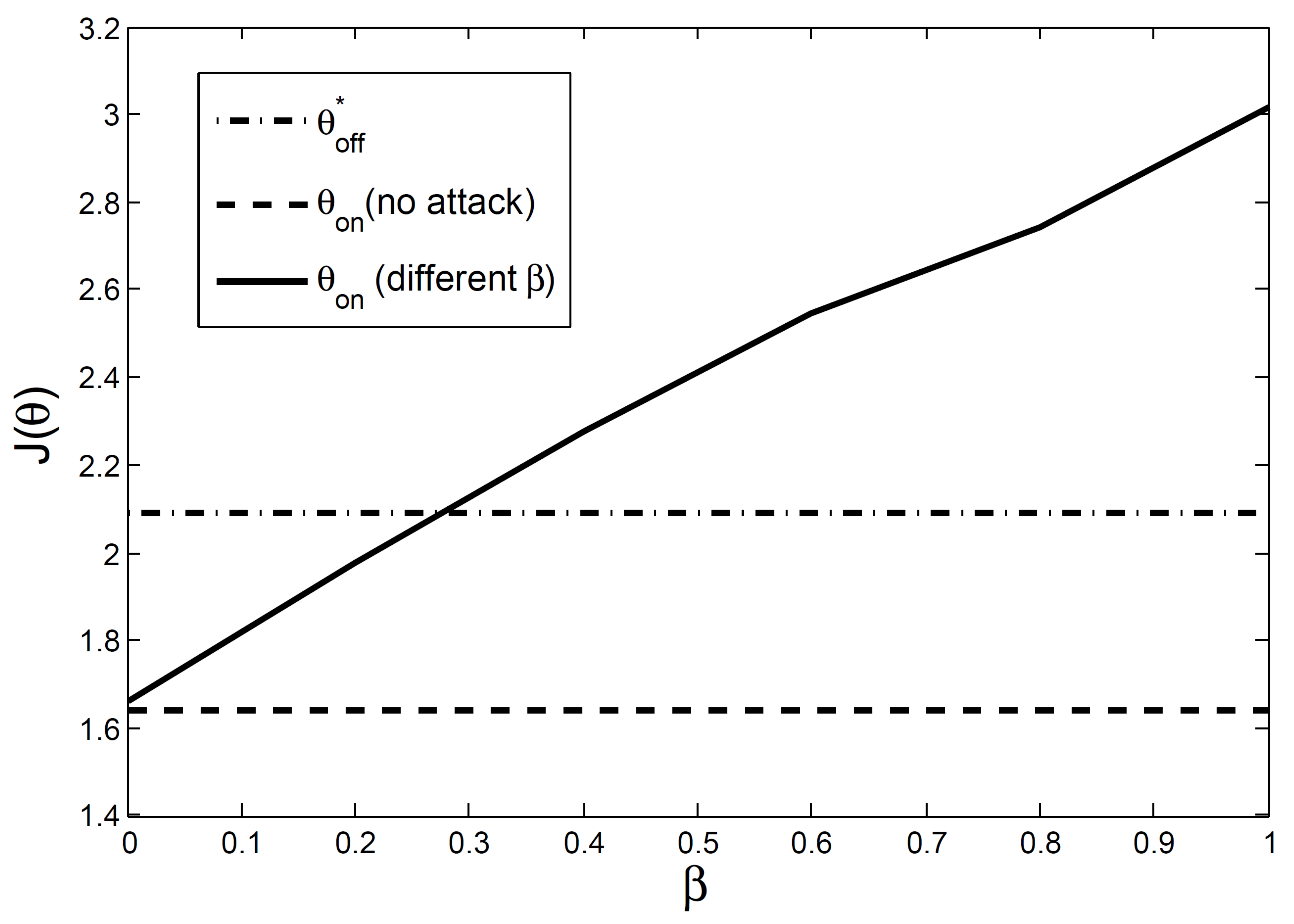}

  \caption{Comparison of $J(\Tilde \theta_{\text{on}})$ under different $\beta$.} \label{fig:simulation_example2}
\end{figure}

Consider a scalar system with parameters $A=1.2$, $C=0.7$, $R=Q=0.8$, $\lambda=0.5$. Suppose that the energy constraint for the sensor is given by $\Psi=\frac{1}{7}\Delta+\frac{6}{7}\delta$. It is easy to verify that the optimal offline schedule is given in a periodic form of $\{1000000, 1000000,...\}$ with performance $J(\theta^\star_{\text{off}})=2.0953$. The parameter for the online schedule is $z_0=2$ with performance $J(\theta_{\text{on}})=1.6399$. As shown in Fig.~\ref{fig:simulation_example1}, when the energy constraint for the attacker is $\beta=\frac{1}{5}$, the estimation performance $J(\Tilde\theta_{\text{on}})$ is better than the performance using the offline schedule $\theta_{\text{off}}^\star$ (thus the sensor may still choose $\theta_{\text{on}}$). However, when the energy budget for the attacker is further increased, e.g., $\beta=\frac{2}{3}$, $J(\Tilde\theta_{\text{on}})$ is worse even than the offline $\theta_{\text{off}}^\star$ (thus the sensor may switch from the online schedule to the offline one to obtain better estimation performance). In Fig.~\ref{fig:simulation_example2}, the comparison of $J(\Tilde\theta_{\text{on}})$ as a function of $\beta$ is provided. As we can see, the approximate value of $\overline \beta$ is about $0.28$.

\section{Conclusion}

In this work, a potential class of malicious attacks against remote state estimation was studied. We proposed a strategy for the attacker which can modify the flag-ACKs from the remote estimator and convey fake information to the sensor. The corresponding effect on the estimation performance is analyzed based on well-established Markov chain theory. The condition for the sensor to use an online schedule instead of an offline schedule is provided. Simulations were provided to illustrate our results. Future works include obtaining the threshold value of $\overline \beta$ analytically, considering other types of fake-ACK attacks and extending the problem into a game-theoretic framework where the attacker need to design their pattern without being detected.

\bibliographystyle{IEEETran}
\bibliography{reference1}

\end{document}